\documentclass[letterpaper]{aa}  
\usepackage{natbib}
\bibpunct{(}{)}{;}{a}{}{,}
\usepackage{graphicx}
\usepackage{txfonts}
\usepackage{epstopdf}
\usepackage{color}
\begin{document}
 \title{
Submillimeter line emission from LMC 30\,Dor: \\ The impact of a
  starburst on a low metallicity environment
}
\titlerunning{Submillimeter line emission from LMC 30\,Dor}
\authorrunning{Pineda, Mizuno,  R\"ollig, Stutzki et al.}

   \author{   
     J.L.\,Pineda\inst{1} \and
     N.\,Mizuno\inst{2} \and
     M.\,R\"ollig\inst{3} \and
 J.\,Stutzki\inst{3} \and
     C.\,Kramer\inst{4} \and
     U.\,Klein\inst{5} \and 
     M.\,Rubio\inst{6} 
          }   

 \offprints{J.\,L.\,Pineda \email{Jorge.Pineda@jpl.nasa.gov}}

   \institute{  Jet Propulsion Laboratory, California Institute of Technology, 4800 Oak Grove Drive, Pasadena, CA 91109-8099, USA
     \and
ALMA-J Project Office, National Astronomical Observatory of Japan, 2-21-1 Osawa, Mitaka, Tokyo 181-8588, Japan
     \and
     KOSMA, I. Physikalisches Institut, Universit\"at zu K\"oln,   
     Z\"ulpicher Stra\ss{}e 77, D-50937 K\"oln, Germany    
\and
Instituto Radioastronom\'{i}a Milim\'{e}trica, Av. Divina Pastora 7, Nucleo Central, 18012 Granada, Spain
\and
  Argelander-Institut f\"ur Astronomie,  Auf dem H\"ugel 71,      D-53121 Bonn, Germany
     \and
     Departamento de Astronom\'{i}a, Universidad de Chile, Casilla 36-D, Santiago, Chile
   }

\date{Received / Accepted } \abstract { The 30 Dor region in the Large
Magellanic Cloud (LMC) is the most vigorous star--forming region in
the Local Group. Star formation in this region is taking place in
low--metallicity molecular gas that is exposed to an extreme
far--ultraviolet (FUV) radiation field powered by the massive compact
star cluster R136.  30\,Dor is therefore ideally suited to study the
conditions in which stars formed at earlier cosmological times.}  {
Observations of (sub)mm and far--infrared (FIR) spectral lines of the
main carbon--carrying species, CO, [C\,{\sc i}] and [C\,{\sc ii}],
which originate in the surface layers of molecular clouds illuminated
by the FUV radiation of young stars, can be used to constrain the
physical and chemical state of the star--forming ISM.  } { We used the
NANTEN2 telescope to obtain high-angular resolution observations of
the $^{12}$CO $J = 4 \to3$, $J = 7\to6$, and $^{13}$CO $ J = 4 \to 3$
rotational lines and [C\,{\sc i}] $^3$P$_1-^3$P$_0$ and
$^3$P$_2-^3$P$_1$ fine-structure submillimeter transitions in
30\,Dor--10, the brightest CO and FIR-emitting cloud at the center of
the 30\,Dor region.  We derived the physical and chemical properties of
the low-metallicity molecular gas using an excitation/radiative
transfer code and found a self-consistent solution of the chemistry and
thermal balance of the gas in the framework of a clumpy cloud PDR
model. We compared the derived properties with those in the N159W
region, which is exposed to a more moderate far-ultraviolet radiation
field compared with 30\,Dor--10, but has similar metallicity.  We also
combined our CO detections with previously observed low--$J$ CO
transitions to derive the CO spectral--line energy distribution in
30\,Dor--10 and N159W.  } { The separate excitation analysis of the
submm CO lines and the neutral carbon fine structure lines shows that
the mid$-J$ CO and [C\,{\sc i}]--emitting gas in the 30\,Dor--10
region has a temperature of about 160\,K and a H$_2$ density of about
10$^{4}$\,cm$^{-3}$.  We find that the molecular gas in 30\,Dor--10 is
warmer and has a lower beam filling factor compared to that of
N159W, which might be a result of the effect of a strong FUV radiation
field heating and disrupting the low--metallicity molecular gas.  We
use a clumpy PDR model (including the [C\,{\sc ii}] line intensity
reported in the literature) to constrain the FUV intensity to about
$\chi_0 \approx 3100$ and an average total H density of the clump
ensemble of about $10^5$\,cm$^{-3}$ in 30\,Dor--10.  }{}

\keywords{astrochemistry -- ISM: globules   -- ISM: molecules -- ISM: individual (30Dor--10)} \maketitle
%

\section{Introduction}

The radiative feedback of star formation on its progenitor molecular
gas is of great interest in modern astrophysics because it can be a
determining factor in the regulation of star formation in galaxies.
In particular, there is special interest in the interaction of stellar
far-ultraviolet (FUV) photons with a lower metallicity, lower
dust-to-gas ratio gas, because it is thought to be the material from
which stars formed at earlier cosmological times.

The region of the interstellar medium where the chemistry and thermal
balance is dominated by the influence of stellar FUV photons is called
a photon-dominated region (or photodissociation region, or PDR;
\citealt{HollenbachTielens99} and references therein). Owing to the
reduced attenuation of FUV photons by dust in low-metallicity PDRs,
the CO molecule is more readily photo-dissociated, increasing the
abundance of C$^{+}$ and C relative to that of CO. Therefore, the
observational signatures of low-metallicity gas (increased [C\,{\sc
ii}]/CO and [C\,{\sc i}]/CO intensity ratios) are a result of the
combined effect of a strong FUV radiation field and of reduced
metallicity. It is therefore difficult to study separately the
influence of these two parameters on the properties of the
low-metallicity molecular gas.

The Large Magellanic Cloud is an ideal extragalactic laboratory in
which to study star formation in low--metallicity environments. Its
proximity \citep[50\,kpc;][]{Feast99} and nearly face-on orientation
\citep[i=35\degr;][]{vanderMarel01} allow us to isolate individual
clouds and study them at the highest spatial resolution possible for
external galaxies. The LMC also shows strongly varying FUV radiation
fields and therefore very different physical conditions in a
low-metallicity \citep[Z$_\odot \simeq 0.4$
Z$_\odot$;][]{Westerlund97} interstellar medium.

Regions of special interest are the molecular clouds associated with
the 30 Doradus H\,{\sc ii} complex. This complex is powered by one of
the most dense and massive star clusters in the local universe,
R136. Therefore, the 30 Doradus region exemplifies the special case
in which the low--metallicity molecular gas is exposed to a very strong
FUV radiation field.   Among several CO-clouds located around
R136, the 30\,Dor--10 cloud \citep{Johansson98} is associated with the
center of the 30\,Dor nebula. The Spitzer 8$\mu$m and CO maps shown in
Figure~\ref{fig:plot_all_spectra} show that this region is the
brightest CO cloud at the center of 30\,Dor and is located at a
projected distance of 20\,pc from the center of R136. This proximity
to R136 and the presence of associated 8$\mu$m sources, which might
indicate embedded star formation, suggest that the gas in 30\,Dor--10
is exposed to an extreme FUV environment. The CO peak of 30\,Dor--10
also associated with the [C\,{\sc ii}] \citep {Poglitsch95} and FIR
\citep{Werner1978} peaks. Although 30\,Dor--10 is the CO peak at the
center of the 30\,Dor nebula, the sub--mm CO and [C\,{\sc i}] line
intensities are weaker compared with other active regions in the LMC
(e.g. N159W; \citealt{Pineda2008}).  
On scales of about 10\,pc, several low--$J$ transitions of $^{12}$CO
and $^{13}$CO have been observed in this cloud
\citep{Johansson98,Minamidani2007,Minamidani2011} as well as the
[C\,{\sc ii}] 158$\mu$m line \citep{Boreiko91,Poglitsch95}. At larger
spatial scales of about 50 pc, the $^{12}$CO $J = 7 \to 6$, $^{12}$CO
$J = 4 \to 3$ and [C\,{\sc i}] $^3{\rm P}_1 \to ^3{\rm P}_0$ lines
have been observed with the AST/RO telescope \citep{Stark97,Kim2006}.

In this paper we present observations of the $^{12}$CO $J = 4 \to$ 3,
 $J = 7 \to$ 6, and $^{13}$CO $J = 4 \to$ 3 rotational and [C\,{\sc
 i}] $^3{\rm P}_1 \to ^3{\rm P}_0$ and $^{3}{\rm P}_{2} \to ^3{\rm
 P}_1$ fine--structure transitions in the 30\,Dor--10 cloud in the LMC
 using the NANTEN2 telescope. The angular resolution of these
 observations allow us for the first time to study submillimeter lines
 at spatial scales of about 10\,pc. We complement this data set with
 low--$J$ $^{12}$CO and $^{13}$CO transitions observed with Mopra, and
 ASTE, with similar angular resolution.


The physical conditions of the molecular gas in 30\,Dor--10 are
 compared with those derived in a region with a similar metallicity but
 with a more moderate incident FUV radiation field, the LMC N159W
 region \citep{Pineda2008}.  N159W is located in one the edges of
 the N159 H\,{\sc ii} region and is actively forming massive stars, as
 suggested by the presence of several embedded massive young stellar
 objects \citep{Chen2010}. However, the FUV and H$\alpha$ fluxes in
 N159 are 10--20 and 7 times lower than in the 30\,Dor region,
 respectively, suggesting that although N159W is an active region it
 does not have the extreme FUV environment to which 30\,Dor--10 is
 exposed.   The comparison between the physical conditions of the gas
 in 30\,Dor--10 and N159W will help us to understand the effects of
 the strength of the FUV field on the properties of the
 low-metallicity molecular gas in the LMC.



This paper is organized as follows. We describe the NANTEN2
observations in Section~\ref{sec:observations} and the observational
results in Section~\ref{sec:observ-results}.  We derive physical
properties of the low-metallicity molecular gas using a radiative
transfer code independently for the CO and [C\,{\sc i}] lines in
Section~\ref{sec:escape-prob-radi}. We give a self-consistent solution
of the chemistry and thermal balance of the gas using a clumpy
photon-dominated region (PDR) model in
Section~\ref{sec:pdr-model-analysis}.
 We compare our results to those
constrained in the LMC--N159W region in
Section~\ref{sec:comparison-with-lmc} and we study the CO spectral
line distribution in both regions in
Section~\ref{sec:co-line-spectral}. We summarize the results in
Section~\ref{sec:summary-conclusions}.

\section{Observations}
\label{sec:observations}

\begin{figure*}[t]
  \centering
  \includegraphics[width=\textwidth,angle=0]{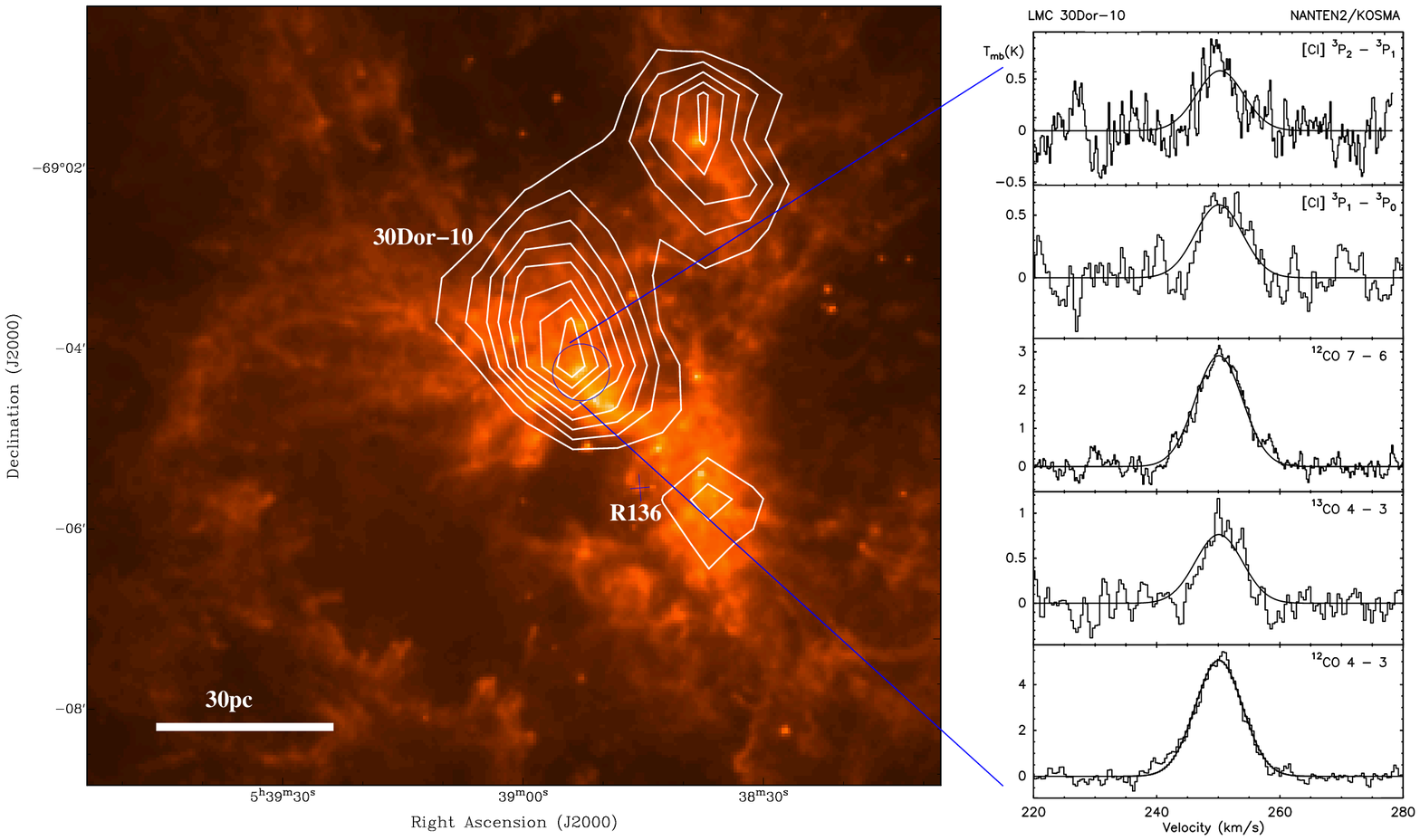}
     \caption{ ({\it Left panel}) Spitzer 8$\mu$m
emission map of the 30\,Dor region \citep{Meixner06} overlaid by an
integrated intensity map of the $^{12}$CO $J = 1 \to 0$ line
\citep{Wong2011}.  The contour levels correspond to 20\% to 80\% of
the peak intensity (11.3 K km/sec), in steps of 10\%. The angular
resolution of the Spitzer 8$\mu$m and CO maps are 2\arcsec\ and
45\arcsec, respectively.  The center of the R136 super star cluster
and the NANTEN2 38\arcsec\ beam are marked by a cross and a circle,
respectively. ({\it Right panel}) Observed lines and corresponding
Gaussian fits.  }
\label{fig:plot_all_spectra}
\end{figure*}

We used the new NANTEN2 4--m telescope situated at 4865\,m altitude at
Pampa la Bola in northern Chile to observe the $^{12}$CO $J = 4 \to 3$
(461.0408\,GHz), $J = 7 \to 6$ (806.6517\,GHz), and $^{13}$CO $J = 4
\to 3$ (440.7654\,GHz) rotational and [C\,{\sc i}] $^3$P$_1-^3$P$_0$
(492.1607\,GHz) and $^3$P$_2-^3$P$_1$ (809.3446\,GHz) fine-structure
transitions toward LMC 30\,Dor--10. The line parameters derived from
the NANTEN2 observations of LMC 30\,Dor--10 are listed in
Table~\ref{tab:gauss} and the observed spectra are shown in
Figure~\ref{fig:plot_all_spectra}.

The observations were made toward the peak $^{12}$CO $J = 1 \to 0$
intensity position in 30\,Dor--10 \citep{Johansson98}, located at
$\alpha$ = 5$^{\tiny \textrm{h}}$38$^{\tiny \textrm{m}}$48\fs56 and
$\delta$ = $-$69\degr04\arcmin43\farcs2 (J2000), using the total-power
observing mode.  We used a reference position at $\alpha$ = 5$^{\tiny
\textrm{h}}$39$^{\tiny \textrm{m}}$33\fs7 and $\delta$ =
$-$69\degr04\arcmin00\farcs0 (J2000), which is free of $^{12}$CO $J =
1 \to 0$ emission.  The duration of each beam switch cycle was varied
between 20 and 30 seconds, depending on the atmospheric stability
during the observations. The final spectra result from total
on--position integration times of 40 min to about 3 hours.  The
pointing was checked regularly on Jupiter, IRC+10216, and IRc2 in
OrionA.  The applied corrections were always smaller than 10\arcsec.
We do not expect significant uncertainties in the line fluxes due
to pointing errors because  the CO flux distribution in 30\,Dor--10 is
smooth at scales smaller than 10\arcsec\,
\citep[e.g.][]{Minamidani2007}.  The angular resolution of the
observations is 38\arcsec\,for the 460--490\,GHz and 26\arcsec\,for
the 810\,GHz channel, respectively.  The main beam efficiencies are
0.5 and 0.4 for the 460--490\,GHz and 810\,GHz channels, respectively.

\begin{table*} [t]
\caption{Gaussian-fit line parameters derived from the NANTEN2
  observations of LMC 30Dor--10 and from the literature}
\label{tab:gauss}
\centering
\begin{tabular}{c c c c c}
\hline\hline
Line & Amplitude$^{(a)}$  & Center$^{(b)}$ & FWHM$^{(b)}$ & Integrated Intensity \\
     &  [K]  &  [km\,s$^{-1}$] & [km\,s$^{-1}$]&[K\,km\,s$^{-1}$]  \\
\hline
\\[-0.2cm]
\multicolumn {5}{l}{NANTEN2 data}
\\

$^{12}$CO $J = 4 \to 3$ & 5.06 (0.19) & 250.1 (0.053) & 9.02 (0.13) & 48.5 (1.9) \\
$^{12}$CO $J = 7 \to 6$ & 2.89 (0.19) & & & 27.7 (1.7) \\
$^{13}$CO $J = 4 \to 3$ & 0.76 (0.14) & & & 7.3 (1.3)  \\

[C\,{\sc i}] $^3$P$_1\to ^3$P$_0$  & 0.58 (0.14) & & & 5.6 (1.3)  \\ 

[C\,{\sc i}] $^3$P$_2\to ^3$P$_1$  & 0.58 (0.19) & & & 5.6 (1.8) \\

\hline
\multicolumn {5}{l}{Literature data$^{(c)}$}\\
$^{12}$CO $J = 1 \to 0$ & 1.1 (0.03) & 250.3 (0.1)   &  8.79 (0.23) & 11.3 (0.27) \\
$^{12}$CO $J = 3 \to 2$ & 5.2 (0.4) & 250.4 (0.1)   &  8.4 (0.23) & 49.6 (0.27) \\
$^{13}$CO $J = 1 \to 0$ & 0.17 (0.02) & 250.7 (0.27) &  5.35 (0.64) & 0.88 (0.12)\\

[C\,{\sc ii}] $^2$P$_{3/2}\to ^2$P$_{1/2}$  & 6.3 (0.6) & 247.8 (1.6) & 20.5 (1.8) & 109.4 (8.5)\\
\hline

\multicolumn {5}{l} {$^{(a)}$ the  rms noise for lines observed with NANTEN2 are for channel widths of 0.37 km s$^{-1}$ and 0.21 km s$^{-1}$   }\\
\multicolumn {5}{l} {for the 460--490 and 810 GHz channels, respectively.}\\
\multicolumn {5}{l}{$^{(b)}$ center and width fixed to $^{12}$CO $J = 4 \to 3$ best--fit
  values for NANTEN2 data other than this line } \\
\multicolumn {5}{l} {$^{(c)}$ references for literature data are presented in  Section 3.2}

\end{tabular}
\end{table*}

\begin{table*} []
\caption{Observed line intensity ratios, corrected for beam coupling
to an assumed common FWHM source size of 90\arcsec }
\label{tab:gauss_corrected}
\centering
\begin{tabular}{c c c c c}
\hline\hline
Species & Ratio$^{(a)}$ & fit error & calibration error &
total error \\
\hline
\\[-0.4cm]
\\
%
%
%
%
%
%
%
%
%
%
$^{12}$CO $J = 7 \to 6$ /$^{12}$CO $J = 4 \to 3$ 
& 0.52 & 0.04 & 28\% & 0.15 \\

[C\,{\sc i}] $^3$P$_2\to ^3$P$_1$  / [C\,{\sc i}] $^3$P$_1\to ^3$P$_0$  
& 0.92 &0.37 &28\%& 0.45 \\

$^{13}$CO $J = 4 \to 3$ / $^{12}$CO $J = 4 \to 3$ 
& 0.15 &0.03 & 17\%& 0.05 \\

[C\,{\sc i}] $^3$P$_2\to ^3$P$_1$  / CO $J = 7 \to 6$ 
& 0.20 &0.07& 17 & 0.09 \\

$^{12}$CO $J=1\to 0$ / $^{12}$CO $J=4\to 3$
& 0.22 &0.05 & 28\% & 0.08 \\



$^{13}$CO $J=1\to 0$ / $^{12}$CO $J=1\to 0$ 
& 0.08 &0.01 & 28\% & 0.02 \\

[C\,{\sc ii}]  $^2$P$_{3/2}\to ^2$P$_{1/2}$/ $^{12}$CO $J=4\to 3$
& 1.24 & 0.13 & 28\%  & 0.37\\

\hline
\multicolumn {5}{l}{$^{(a)}$  ratio of amplitudes from Gaussian fits} \\
\end{tabular}
\end{table*}

The observations were conducted with a dual-channel 460/810\,GHz
receiver installed for verifying the submillimiter performance of the
telescope.  Double-sideband (DSB) receiver temperatures were $\sim$250\,K in the lower channel and $\sim$ 750\,K in the upper channel. The
intermediate frequencies (IF) are 4\,GHz and 1.5\,GHz,
respectively. The latter IF allows simultaneous observations of the
$^{12}$CO $J = 7 \to 6$ line in the lower and of the [C\,{\sc i}]
$^3$P$_2-^3$P$_1$ line in the upper sideband. These two lines were
observed simultaneously with one of the lines in the 460\,GHz
channel. As backends we used two acousto-optical spectrometers (AOS)
with a bandwidth of 1\,GHz and a channel resolution of 0.37\,km
s$^{-1}$ at 460\,GHz and 0.21\,km s$^{-1}$ at 806\,GHz.

As discussed in \citet{Pineda2008}, the main sources of uncertainty in
the absolute line calibration are the precision with which the beam
efficiencies are determined and the accuracy of the atmospheric
calibration. We estimate that both sources of uncertainty contribute
to a total absolute calibration uncertainty of 20\%. In our analysis
we added these errors quadratically with the radiometric noise,
assumed to be given by the formal errors of the Gaussian
fits. Propagating these absolute calibration uncertainties for line
ratios between lines in the 460--490 and 810\,GHz channels, we obtain
a relative calibration uncertainty of 28\%.  Line ratios between lines
located in the same frequency channel do not suffer from the
uncertainty on the determination of the beam efficiencies ($\sim$10\%)
and we propagate the errors only considering the uncertainties on the
atmospheric calibration ($\sim$17\%).


\section{Observational results}
\label{sec:observ-results}
\subsection{NANTEN2 data}
\label{sec:nanten2-data}


The spectra observed toward 30\,Dor--10 are shown in
Fig.~\ref{fig:plot_all_spectra}, together with Gaussian fits whose
results are listed in Table~\ref{tab:gauss}. We determined the line
center and width from the high signal-to-noise $^{12}$CO $J = 4 \to 3$
spectrum and we held them fixed for the Gaussian fit of the line
amplitude of the other lines. This method is adequate because their
profiles are, within their limited signal-to-noise ratio, consistent
with that of the $^{12}$CO $J = 4 \to 3$ line.

To calculate intensity ratios between lines in the upper and lower
frequency channels we need to account for the different beam sizes
(38\arcsec and 26\arcsec, respectively).  We therefore determined the
source extent using lower resolution data. As described in
\citet{Pineda2008}, assuming a Gaussian source distribution with FWHM
$\Theta_s$ and a source peak brightness temperature of $T_{s,peak}$,
the beam filling correction gives a main beam brightness temperature
in a beam of FWHM $\Theta_b$ of $
T_{mb}={\Theta_s^2}/{(\Theta_s^2+\Theta_b^2}) T_{s,peak}$ .  The ratio
of the main beam brightness temperature in two different beams 1 and 2
is given by $R_{1,2} = {T_{mb,1}}/ {T_{mb,2}} = {(\Theta_{b,2}^2 +
\Theta_s^2)} /{(\Theta_{b,1}^2+\Theta_s^2)} $.  With this, we can
derive the source size from the observed brightness ratio in two beam
sizes:
\begin{equation}\label{eq:1}
\Theta_s^2 = \frac {R_{1,2} \Theta_{b,1}^2 - \Theta_{b,2}^2} {1-R_{1,2}}.
\end{equation}
With the source size we can correct the observed intensity ratios for
the beam coupling of the different beam sizes at two observing bands
$x$ and $y$ to the source intrinsic intensity ratio $R_s$ via

\begin{equation}\label{eq:2}
R_s = \frac {T_{mb,x}}
{T_{mb,y}
} \, \frac {\Theta_s^2+\Theta_{b,x}^2}
{\Theta_s^2+\Theta_{b,y}^2} .
\end{equation}
We used observations of the [C\,{\sc i}] $^3$P$_1\to ^3$P$_0$ line
made with the SWAS satellite (Frank Bensch, priv. comm.) with an
angular resolution of 258\arcsec. The integrated line intensity of
this line observed by SWAS is 0.44 K km\,s$^{-1}$. With NANTEN2 we
obtained an integrated line intensity of 5.58 K km\,s$^{-1}$ in a
38\arcsec\,beam. Thus, from Equation~(\ref{eq:1}) we obtain a source
size of $\Theta_s$=64\arcsec, which according to
Equation~(\ref{eq:2}), corresponds to a 18\% correction to the line
ratios between the upper and lower channels. We also determined the
source size by considering the $^{12}$CO $J = 1 \to 0$ integrated line
intensity observed by the NANTEN telescope \citep[e.g.][]{Fukui2008}
of 4.1 K km\,s$^{-1}$ in a 158\arcsec\,beam and by the Mopra telescope
(J\"urgen Ott, priv. comm.)  of 11.3 K km\,s$^{-1}$ in a
33\arcsec\,beam. Equation~(\ref{eq:1}) results in a source size of
$\Theta_s$=110\arcsec, corresponding to a 8\% correction. We assumed
an intermediate value for the source size of
$\Theta_s$=90\arcsec\,corresponding to a correction of 10\%.  A source
size of 90\arcsec\ is consistent with the angular size of
30\,Dor--10's CO--emitting region shown in
Figure~\ref{fig:plot_all_spectra}.  We note that assuming a source
size introduces an uncertainty of about 5\% to the line ratios between
the upper and lower channels.  The correction also assumes that the
source structure is the same at the spatial scales observed with the
460 and 810 GHz channels.  The validity of this assumption likely
dominates the uncertainties in the line ratios.  However, as we will
see in Sections~\ref{sec:escape-prob-radi} and
\ref{sec:pdr-model-analysis}, the good fit obtained for the physical
parameters supports this assumption.  Table~\ref{tab:gauss_corrected}
shows the line ratios thus corrected, which we used for the excitation
analysis discussed below.

\citet{Kim2005} presented maps of the $^{12}$CO $J = 4 \to 3$ line in
the 30\,Dor region with a 109\arcsec\, beam using the AST/RO
telescope. In addition to the $^{12}$CO $J = 4 \to 3$ component at
$\sim$250\,km\,s$^{-1}$ they detected another component at
$\sim$302\,km\,s$^{-1}$, a factor of $\sim$3 weaker. We did not find
this additional component in our $^{12}$CO $J = 4 \to 3$ spectrum.
Considering their derived FWHM line--width for this component of
5.1\,km\,s$^{-1}$, we derive an 3$\sigma$ upper limit of 0.52
K\,km\,s$^{-1}$, which is a factor of $\sim$10 weaker than the main
component of $^{12}$CO $J = 4 \to 3$ in 30\,Dor--10.

\subsection{Additional lines from the literature }
\label{sec:addit-lines-from}

To better constrain the physical parameters of the emitting gas in
30\,Dor--10, we combined the lines observed with NANTEN2 with
additional observations of the $^{12}$CO $J=1\to 0$ , $J=3\to 2$,
$^{13}$CO $J=1\to 0$ and [C\,{\sc ii}] 158 $\mu$m lines.  The
$^{12}$CO and $^{13}$CO $J=1\to 0$ lines were observed with the ATNF
Mopra 22\,m telescope (J\"urgen Ott, priv. comm.)  at the position of
the NANTEN2 observations. The data have an angular resolution of
33\arcsec\, and a channel width of 0.7\,km\,s$^{-1}$.  The $^{12}$CO
$J=3\to 2$ observations were observed with the ASTE telescope and were
presented by \citet{Minamidani2007}. The angular resolution for both
lines is about 22\arcsec and the channel width is 0.45 km s$^{-1}$.
The Kuiper Airborne Observatory (KAO) observed velocity--resolved
[C\,{\sc ii}] 158 $\mu$m emission toward 30\,Dor--10 with an angular
resolution of 43\arcsec \citep{Boreiko91}.  The pointing of the KAO
observations is about $20$\arcsec\,farther southwest than the NANTEN2
position.  We adopted their values without further
corrections. Indeed, the integrated intensity of the [C\,{\sc ii}]
line from the KAO observations of $7.7 \times 10^{-4}$ erg cm$^{-2}$
s$^{-1}$ sr$^{-1}$ is consistent to within 30\% with the integrated
intensity at the position of the NANTEN2 observations in the [C\,{\sc
ii}] map published by \cite{Poglitsch95}, observed with low--velocity
resolution ($\Delta$v = 75 km s$^{-1}$) using the FIFI spectral
imaging instrument on the KAO.  In Table~\ref{tab:line_parameters} we
present a summary of rest frequencies, energies above the ground
level, and critical densities of the spectral lines used in our
analysis.

We assumed that the uncertainty of the absolute calibration of all
independently observed lines is about 20\%, and propagated these
errors to obtain a line ratio calibration accuracy of 28\%.
Table~\ref{tab:gauss_corrected} quotes the derived total errors,
obtained by quadratically summing the formal fit error and the
calibration uncertainty.

\begin{figure}[t]
  \centering
  \includegraphics[width=0.34\textwidth,angle=0]{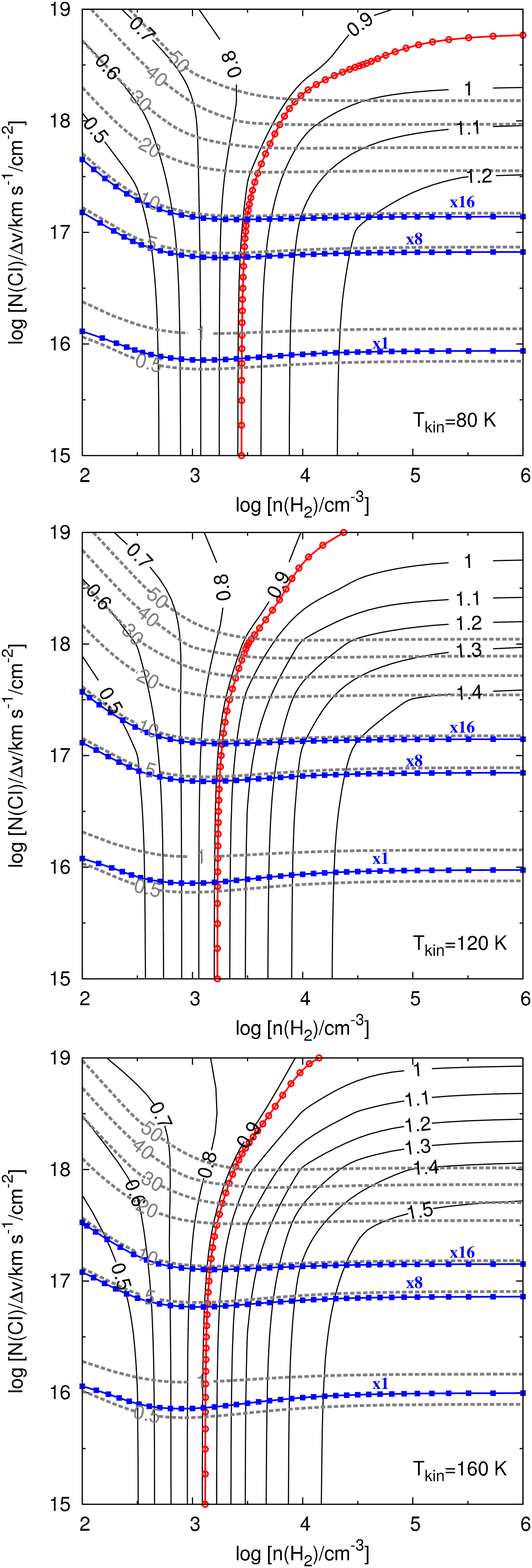}
     \caption{Comparison between the observations and the predictions
       from the radiative transfer model for the [C\,{\sc i}]
       $^3$P$_2\to ^3$P$_1$/[C\,{\sc i}] $^3$P$_1\to ^3$P$_0$ line
       ratio (black--solid lines) and the [C\,{\sc i}] $^3$P$_1\to
       ^3$P$_0$ absolute intensity (gray--dashed lines) for kinetic
       temperatures of 80, 120, and 160\,K.  The observed [C\,{\sc i}]
       $^3$P$_2\to ^3$P$_1$/[C\,{\sc i}] $^3$P$_1\to ^3$P$_0$ line
       ratio of 0.92 is shown as a red line with circles. The observed
       [C\,{\sc i}] $^3$P$_1\to ^3$P$_0$ line intensity of 0.58\,K and
       its value scaled by a beam filling factor of 1/8 and 1/16 are
       shown as blue lines with boxes. }
\label{fig:ci_ratio}
\end{figure}

\begin{figure}[t]
  \centering
  \includegraphics[width=0.34\textwidth,angle=0]
{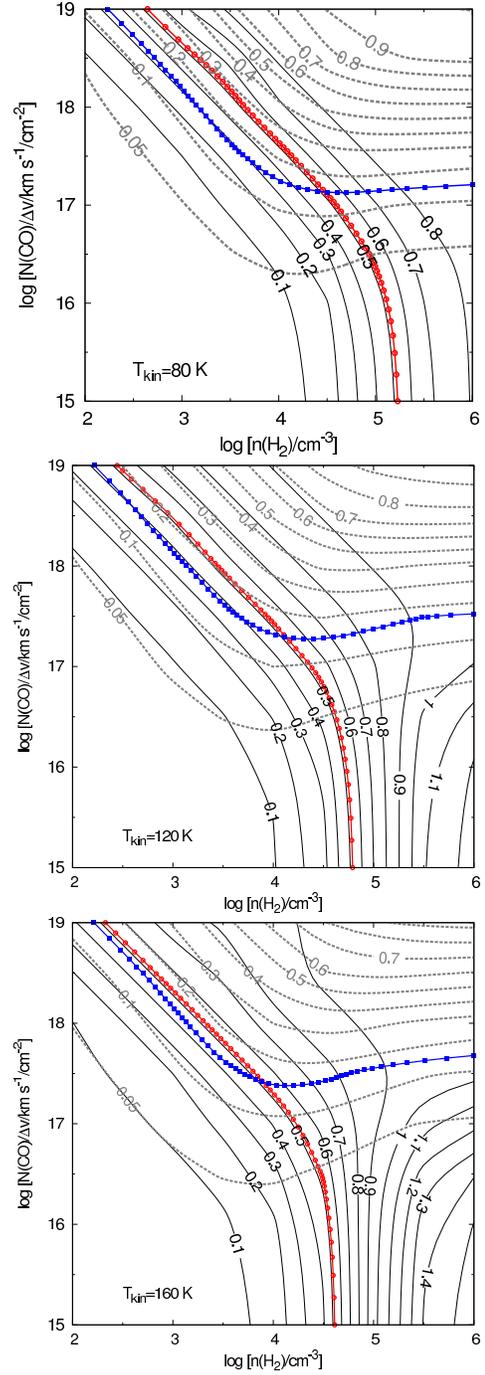}
     \caption{Comparison between the observations and the predictions
       from the radiative transfer model for the $^{12}$CO $J = 7 \to
       6$ to $J = 4 \to 3$ (black--solid contours) and $^{13}$CO to $^{12}$CO
       $J =4 \to 3$ (gray--dashed contours) line ratios for kinetic
       temperatures of 80, 120, and 160\,K. The plots assume a
       fractional abundance of $^{12}$CO/$^{13}$CO of 35. The observed
       $^{12}$CO $J = 7 \to 6$ to $J = 4 \to 3$ ratio of 0.52 is shown
       as a blue line with boxes while the $^{13}$CO to $^{12}$CO $J =4 \to 3$
       ratio of 0.15 is shown as a red line with circles.}
\label{fig:co_ratios}
\end{figure}

\begin{figure}[t]
  \centering
  \includegraphics[width=0.45\textwidth,angle=0]
{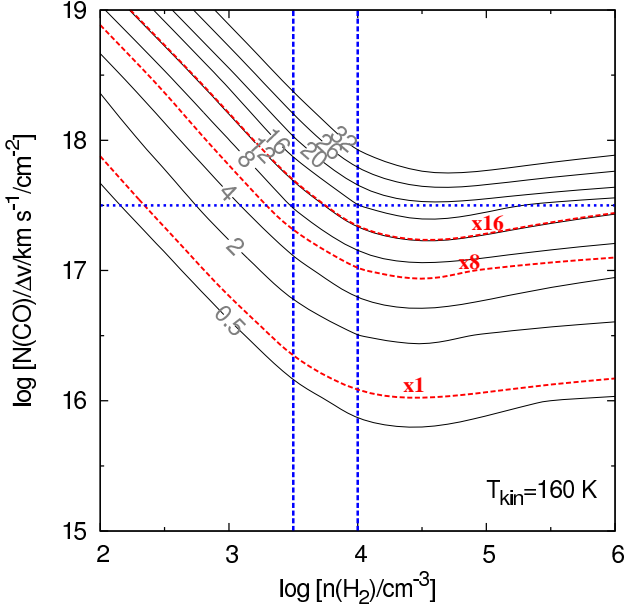}
     \caption{Radiative transfer predictions and the observed absolute
       intensity of $^{13}$CO $J=4\to 3$ for $T_{\rm kin} = 160$\,K.
       The three red--dashed contours show the observed intensity and
       its value scaled up by factors of 8 and 16, respectively. The
       blue--dashed lines show the range of column and volume
       densities constrained by the analysis of the CO line
       ratios. The absolute intensity of $^{13}$CO $J=4\to 3$ needs to
       be scaled up by a factor of about 16 to reproduce the
       constrained column and volume densities.}
\label{fig:13co_abs}
\end{figure}

\begin{figure}[t]
  \centering
  \includegraphics[width=0.45\textwidth,angle=0]
{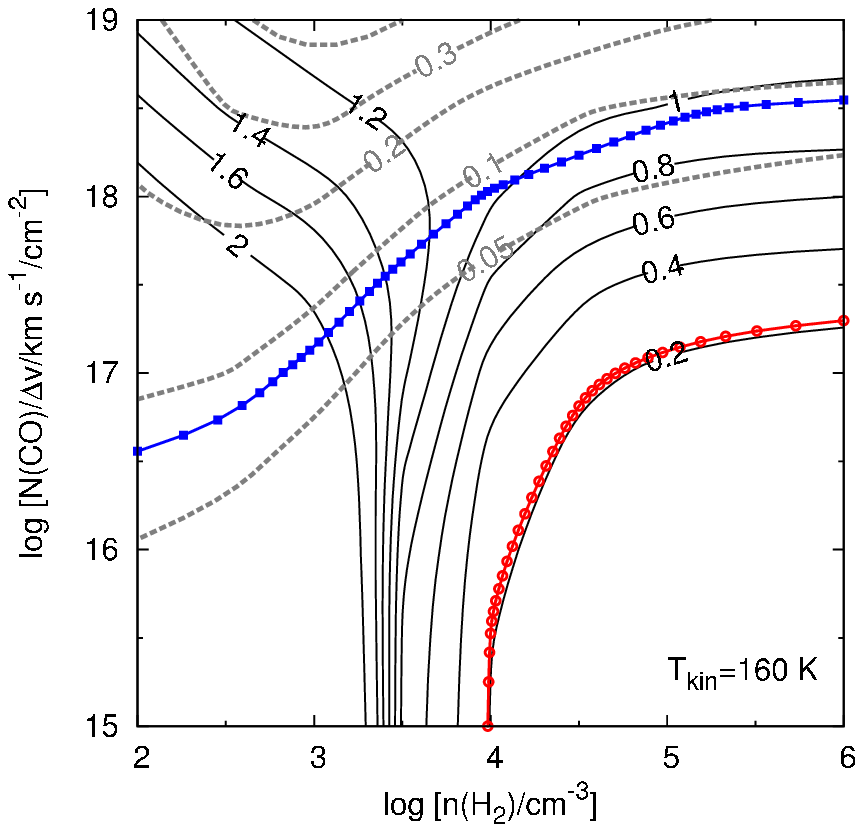}
     \caption{Comparison between the observations and the predictions
       from the radiative transfer model for the $^{12}$CO $J = 1 \to
       0$ and $J = 4 \to 3$ line ratios (black--solid  contour) and the
       $^{13}$CO to $^{12}$CO $J=1\to 0$ line ratio (gray--dashed
       contour) for $T_{\rm kin} = 160$\,K.  The observed $^{13}$CO to
       $^{12}$CO $J =1 \to 0$ ratio of 0.08 is shown as a blue line with boxes
       while the $^{12}$CO $J = 1 \to 0$ to $J = 4 \to 3$ ratio of
       0.22 is shown as a red line with circles. }
\label{fig:co_ratio}
\end{figure}

\section{Excitation analysis}
\label{sec:escape-prob-radi}

We first constrained the properties of the emitting gas by comparing
line intensity ratios and absolute intensities with the results of a
radiative transfer code. We analyzed [C\,{\sc i}] and CO separately,
but assume that they originate from the same regions with identical
beam filling factors. This assumption is justified considering that
the CO and [C\,{\sc i}] emission arise from the FUV illuminated
surfaces of clouds that are unresolved at the scale of our
observations.

We used the RADEX non--LTE radiative transfer code by
\citet{vanderTak2007}, using the uniform sphere approximation, to
calculate line intensities as a function of the kinetic temperature,
H$_2$ volume density, and the column density per velocity interval
$N$/$\Delta$v of the species of interest. The collision cross sections
were taken from the Leiden Atomic and Molecular Database
\citep[LAMDA;][]{Schoeier2005}.  The code provides the average line
intensity of a spherical cloud. We therefore scaled the data by a
filling factor to compare them with the observed absolute intensities.
As mentioned above, considering a medium consisting of clumps that are
much smaller than the scale of our observations, the C\,{\sc i} and CO
species can be considered as occupying the same volume, hence the line
ratios are independent of beam dilution. For optically thin clumps (or
for clumps with moderate opacity), the observed--to--model integrated
intensity ratio gives the ratio of the beam--averaged column density
to the clump intrinsic column density, assuming that the total
integrated intensity is proportional to the number of clumps, i.e. no
velocity crowding is present. We therefore determined the beam filling
by comparing the observed absolute intensity of a line with a
constrained model.

\begin{table} [t]
\caption{Parameters for different relevant molecular and atomic
transitions. The listed critical densities were calculated for
collisions with $p$--H$_2$ and a kinetic temperature of 100K. The
critical densities were calculated using the Einstein A-- and
collisional rate coefficients taken from the Leiden Atomic and
Molecular Database \citep[LAMDA;][]{Schoeier2005}.  }
\label{tab:line_parameters}
\centering
\begin{tabular}{c c  c c }
\hline\hline
Line   & Frequency/GHz  & $E_{\rm upper}$/K & $n_{\rm cr}/{\rm cm}^{-3}$ \\
\hline
\\
$^{12}$CO $J = 1\to 0$   &  115.2712  & 5.5  &   2$\times$10$^{3}$ \\
						 		                   
$^{12}$CO $J = 3 \to 2$  & 345.7959   & 33.2  &  4$\times$10$^{4}$ \\
						 		                   
$^{12}$CO $J = 4 \to 3$  & 461.0408   & 55.3  &  9$\times$10$^{4}$ \\
						 		                   
$^{12}$CO $J = 7 \to 6$  & 806.6517   & 154.9  & 4$\times$10$^{5}$  \\ 
										                   
$^{12}$CO $J = 9 \to 8$  & 1036.912   &  248.9  &  1$\times$10$^{6}$ \\

$^{13}$CO $J=  1 \to 0$  & 110.2013  & 5.3 & 2$\times$10$^{3}$ \\
			
$^{13}$CO $J = 4 \to 3$                     & 440.7654  & 52.9 & 8$\times$10$^{4}$ \\
					  
[C\,{\sc i}] $^3$P$_1\to ^3$P$_0$           & 492.1607  & 23.6 & 1$\times$10$^{3}$ \\
								  
[C\,{\sc i}] $^3$P$_2\to ^3$P$_1$           & 809.3446  & 62.5 & 1$\times$10$^{3}$   \\

[O\,{\sc i}] $^3$P$_1\to ^3$P$_2$           & 4744.777 &  227.7  & 6$\times$10$^{5}$  \\
			
[O\,{\sc i}] $^3$P$_0\to ^3$P$_1$           & 2060.069 &  326.6  & 7$\times$10$^{4}$ \\
			
[C\,{\sc ii}]  $^2$P$_{3/2}\to ^2$P$_{1/2}$ & 1900.5377  & 91.2 & 5$\times$10$^{3}$ \\
\hline
\end{tabular}
\end{table}

\subsection{[CI]-emission}
\label{sec:ci-emission}

In Table~\ref{tab:gauss_corrected} we see that the ratio between the
two [C\,{\sc i}] lines is 0.92. We compare this ratio with predictions
from the radiative transfer calculations in
Figure~\ref{fig:ci_ratio}. We present model line ratios for three
different kinetic temperatures 80, 120, and 160\,K, as a function of
the [C\,{\sc i}] column density per velocity interval, $N_{\rm
[C\,{{\sc I}]}}/\Delta {\rm v}$, and H$_2$ volume density, $n({\rm
H}_2)$. The coverage in $N_{\rm [C\,{{\sc I}]}}/\Delta {\rm v}$ is
$10^{15}-10^{19}$\,cm$^{-2}$ and in $n({\rm H}_2)$ is
$10^{2}-10^{6}$\, cm$^{-3}$. The observed ratio can only be reproduced
by kinetic temperatures above 60\,K and H$_2$ densities of about $1-3
\times 10^{3}$\,cm$^{-3}$. This H$_2$ density is similar to, or
somewhat higher than, the critical density of the [C\,{\sc i}]
$^3$P$_1\to ^3$P$_0$ line ($\sim 1\times 10^3$\,cm$^{-3}$), and
therefore its emission is close to being thermalized. From the
analysis of the CO emission (see below) we found that the
beam--filling factor in 30\,Dor--10 is about 1/16, so that the
observed [C\,{\sc i}] $^3{\rm P}_1-^3{\rm P}_0$ brightness of 0.58\,K
should be scaled up to 9.3\,K. This brightness temperature corresponds
to a [C\,{\sc i}] column density per velocity interval of about
$1\times10^{17}$ cm$^{-2}$/km s$^{-1}$. The multiplication by a
filling factor is appropriate because, in the range of kinetic
temperatures and H$_2$ densities inferred above, the [C\,{\sc i}]
$^3{\rm P}_1-^3{\rm P}_0$ line is likely optically thin.

\subsection{CO-emission}
\label{sec:co-emission}

We compare the observed $^{12}$CO $J=7\to6/^{12}$CO $J=4\to3$ and
$^{13}$CO $J=4\to3/^{12}$CO $J=4\to3$ ratios with predictions of the
radiative transfer code in Figure~\ref{fig:co_ratios}. We assumed the
$^{12}$CO/$^{13}$CO abundance ratio of 35 derived in 30\,Dor--10 by
\citet{Heikkilae99}.  We required that the volume density of the
CO--emitting region is equal to, or somewhat higher than, that
constrained for the [C\,{\sc i}] emission, because we expect that both
[C\,{\sc i}] and mid--J CO lines are emitted from adjacent layers in a
PDR--like structure\footnote{For an example of a typical temperature
distribution and abundance structure of the main carbon species in
PDRs, see e.g. Figures 7 and 9 in \citealt{Tielens1985}.}, where the
neutral carbon is located in an equal or lower--density region
compared to the one at which CO is located, and that the cloud's
volume density profile is smooth. In Figure~\ref{fig:co_ratios}, this
condition is only valid for kinetic temperatures higher than 120K for
a volume density range between $3\times10^{3}-10^{4}$\,cm$^{-3}$.  The
intersection of the two CO line ratios constrains the CO column
density per velocity interval to about $3\times10^{17}$\,cm$^{-2}$/km
s$^{-1}$ at $T_{\rm kin} \simeq 160$\,K.  The derived kinetic
temperature and H$_2$ density are consistent with previously published
excitation analyzes based on CO observations
\citep{Johansson98,Kim2006,Minamidani2007,Minamidani2011}.
%
%
%

Figure~\ref{fig:13co_abs} shows that the model--predicted
clump--averaged absolute line intensity of the $^{13}$CO $J=4\to 3$
line, at the column and volume density regime constrained by the
observed line ratios, is about 16 times higher than that observed at
$T_{\rm kin} =$160\,K.  This implies a beam filling factor for CO of
about 1/16, which corresponds to a typical clump size of 10\arcsec.
Recent ATCA observations of HCO$^+$ revealed several compact clumps in
30\,Dor--10 (Anderson et al. in preparation). The clumps have a
typical angular size of 10\arcsec, and are marginally resolved at the
resolution of their observations, 6\arcsec. More observations with
higher angular resolution are required to confirm the observed clump
sizes in 30\,Dor--10.

In Figure~\ref{fig:co_ratio} we consider line ratios involving the
$^{12}$CO and $^{13}$CO $J = 1 \to 0$ transitions for $T_{\rm kin}
=$160\,K.  The $^{12}$CO $J = 1 \to 0$ to $J = 4 \to 3$ ratio of 0.21,
however, suggests lower column and higher volume densities than those
constrained above. The low $^{12}$CO $J = 1 \to 0$ to $J = 4 \to 3$
ratio is due to the observed $^{12}$CO $J = 1 \to 0$ line that is too
weak for the physical conditions we derived.  The observed
$^{13}$CO/$^{12}$CO ratio is therefore also affected by this weaker
$^{12}$CO $J = 1 \to 0$ emission. This may indicate an additional
colder gas component. We were only able to reproduce absolute
intensities of the $^{12}$CO and $^{13}$CO $J = 1 \to 0$ lines using a
model with the conditions obtained for the column and volume density
derived for [C\,{\sc i}] (Section~\ref{sec:ci-emission}) and kinetic
temperatures of $T_{\rm kin}$=10--30\,K.  This is consistent with
observations of higher density tracers, suggesting a colder
($\sim$50\,K) and denser ($\sim10^5$ cm$^{-3}$) gas component in
30\,Dor--10 \citep{Heikkilae99}.

%
%


\subsection{[C\,{\sc i}]  and CO column densities}
\label{sec:column_densities}

Combining our estimate of the CO and C\,{\sc i} column density per
velocity interval, the observed line width of 9 km s$^{-1}$, and the
beam filling factor of 1/16, we derive a beam--averaged column density
of 1.7$\times10^{17}$ cm$^{-2}$ and 5.6$\times10^{16}$ cm$^{-2}$ for
CO and C\,{\sc i}, respectively.  Thus, the ratio of the column
density of [C\,{\sc i}] to that of CO is $\sim$3.  Since the $^{12}$CO
and $^{13}$CO $J = 1\to 0$ lines indicate an additional, cooler
component, this ratio is an upper limit.  Assuming that all gas--phase
carbon is in the form of C\,{\sc i} and CO in their respective
line--emitting regions, we used a carbon abundance relative to H$_2$
in the LMC of $1.6 \times 10^{-4}$ (derived from [C]/[H]=8$\times
10^{-5}$; \citealt{Dufour1982}) to obtain an H$_2$ mass of
1625\,M$_\odot$ associated with CO and of 531\,M$_\odot$ associated
with C\,{\sc i}.  Thus, the total mass within the beam of our
observations is 2156\,M$_\odot$.  Note that our beam size of
38\arcsec\,FWHM corresponds to 9.5\,pc at the distance of the LMC and
hence covers 1/2 of the 30\,Dor--10 cloud's extent of about 20 pc. If
the H$_2$ column density is uniform over 30\,Dor--10, the total mass
of the cloud should be a factor of 4 larger than the mass derived
within the beam of our observations, reflecting the difference in
area. Thus, the total H$_2$ mass in 30\,Dor--10 should be about
8.6$\times 10^4$ M$_\odot$.  This derived mass agrees well with the
CO-luminosity mass of $8.5 \times 10^4$ M$_\odot$ derived by
\cite{Johansson98} assuming a CO--to--H$_2$ conversion factor of
4$\times10^{20}$ cm$^{-2}$ (K km s$^{-1}$)$^{-1}$
\citep[e.g.][]{Israel97,Hughes2010,Pineda2009}.  Note that some H$_2$
might be not traced by CO or [C\,{\sc i}] but by [C\,{\sc ii}] (the
``hidden H$_2$ gas''). This can account for about 30--50\% of the
H$_2$ in the Galaxy
\citep[e.g][]{Grenier2005,Langer2010,Velusamy2010,Bernard2011}, and is
expected to have an even larger mass in low--metallicity environments
\citep{Madden1997,Wolfire2010}.  Therefore, the mass estimate derived
here should be considered a lower limit to the total H$_2$ mass.

\section{PDR-model analysis}
\label{sec:pdr-model-analysis}

\begin{table} [t]
\caption{Comparison between observed line integrated intensities and
the predictions from the KOSMA$-\tau$ PDR model and the RADEX
radiative transfer code. All intensities are in units of erg cm$^{-2}$
s$^{-1}$ sr$^{-1}$. The observed and RADEX intensities are corrected
for beam dilution. The PDR model absolute intensities are calculated
assuming a common FWHM source size of 90\arcsec.  }
\label{tab:model_comparison}
\centering
\begin{tabular}{c c c c }
\hline\hline
Line   & Observed & KOSMA$-\tau^{a}$ & RADEX$^{b}$ \\
\hline
\\
$^{12}$CO $J = 1\to 0$   & 2.8$\times 10^{-7}$(0.2)   & 8.7$\times 10^{-7}$  & 8.1$\times 10^{-7}$  \\
										                   
$^{12}$CO $J = 3 \to 2$  & 3.0$\times 10^{-5}$(0.3)   & 3.4$\times 10^{-5}$  & 3.2$\times 10^{-5}$ \\
										                   
$^{12}$CO $J = 4 \to 3$  & 7.8$\times 10^{-5}$(1.6)   & 7.8$\times 10^{-5}$  & 6.8$\times 10^{-5}$  \\
										                   
$^{12}$CO $J = 7 \to 6$  & 2.2$\times 10^{-4}$(0.4)   & 1.9$\times 10^{-4}$  & 2.1$\times 10^{-4}$ \\ 
										                   
$^{12}$CO $J = 9 \to 8$  &  $<$3.3$\times 10^{-4}$     & 9.4$\times 10^{-5}$  & 1.9$\times 10^{-4}$ \\

$^{13}$CO $J=  1 \to 0$  & 1.9$\times 10^{-8}$(0.3)  & 6.8$\times 10^{-8}$ & 3.1$\times 10^{-8}$ \\
			
$^{13}$CO $J = 4 \to 3$                     & 1.0$\times 10^{-5}$(0.3)  & 1.1$\times 10^{-5}$ & 1.0$\times 10^{-5}$ \\
			
[C\,{\sc i}] $^3$P$_1\to ^3$P$_0$           & 1.1$\times 10^{-5}$(0.3)  & 1.0$\times 10^{-5}$ & 5.5$\times 10^{-6}$ \\
			
[C\,{\sc i}] $^3$P$_2\to ^3$P$_1$           & 4.1$\times 10^{-5}$(1.7)  & 4.4$\times 10^{-5}$ & 3.5$\times 10^{-5}$ \\
			
[C\,{\sc ii}]  $^2$P$_{3/2}\to ^2$P$_{1/2}$ & 1.2$\times 10^{-2}$(0.01) & 2.4$\times 10^{-2}$ & 4.5$\times 10^{-3}$ \\
\hline
\multicolumn{4}{l}{(a) Best--fit model: $n_{\rm ens} =3.3\times10^5$\,cm$^{-3}$, $M_{\rm ens} = 1.2\times10^5$ M$_{\odot}$,}\\
\multicolumn{1}{l}{$\chi_0 = 3160$.  }\\
\multicolumn{4}{l}{(b) Best--fit model: $n_{\rm H_2}=10^4$\,cm$^{-3}$, $T_{kin}=160$\,K, beam filling 1/16,} \\
\multicolumn{4}{l}{$N_{\rm CO}=1.7\times10^{17}$\,cm$^{-2}$,$N_{\rm C^0}=5.6\times10^{16}$\,cm$^{-2}$. } 
\end{tabular}
\end{table}

In the following we investigate whether our observations toward
30\,Dor-10 can be explained in terms of an ensemble of PDRs
distributed in a clumpy interstellar medium. We used the KOSMA--$\tau$
PDR model \citep{Stoerzer96,Roellig06} to model an ensemble of
spherical clumps with a power--law mass spectrum $dN/dM\propto
M^{-1.8}$ and a mass-size relation $M \propto r^{2.3}$
\citep{Cubick2008}.  The model provides a self--consistent solution of
the chemistry and thermal balance.  The free parameters of the clumpy
PDR ensemble are (1) average ensemble total H volume density ($n_{\rm
H}$+2$n_{\rm H_2}$), $n_{\rm ens}$, (2) ensemble mass, $M_{\rm ens}$,
(3) strength of the FUV radiation field in units of the
\citet{Draine78} field, $\chi_0$, and (4) the minimum and maximum mass
of the clump ensemble, [$M_{\rm min},M_{\rm max}$]. We fix the
metallicity in the model to Z$_\odot=0.4$. We attempted to fit the
observed lines with a model with Z$_\odot=1$, but this solar
metallicity model could not reproduce the mid--$J$ CO observations
with a $\chi^2$ about factor of 10 larger than for the
low--metallicity model.  Note that we assumed that the clumps do not
spatially overlap, therefore we considered optical depth effects only
within individual clumps.  We also assumed that the FUV field is
uniform within the beam of the observations.

We constrained the model with the absolute intensities of all five
lines observed with NANTEN2 together with the $^{12}$CO $J = 3 \to 2$
and [C\,{\sc ii}] 158$\mu$m lines taken from the literature
(Section~\ref{sec:addit-lines-from}).  Fitting absolute line
intensities instead of line ratios allowed us to study the internal
structure of the cloud (clumpiness) via the area and volume filling
factors, and it adds the cloud mass and size as model parameters. The
source size was fixed to be 90$\arcsec$
(Section~\ref{sec:nanten2-data}).  We did not include the $^{12}$CO
and $^{13}$CO $J = 1 \to 0$ transitions in the model fit because these
transitions are likely affected by optical depth effects between
clumps that are not accounted for by the model. We still compared the
intensities of the $^{12}$CO and $^{13}$CO $J = 1 \to 0$ transitions
with the predictions of the PDR model and discuss below the possible
reason of the discrepancy between model and observations.  We used
simulated annealing to find the optimum model parameter combination
that predicts absolute line intensities that match those observed.


\begin{figure}[h]
  \centering
  \includegraphics[width=0.465\textwidth,angle=0]
{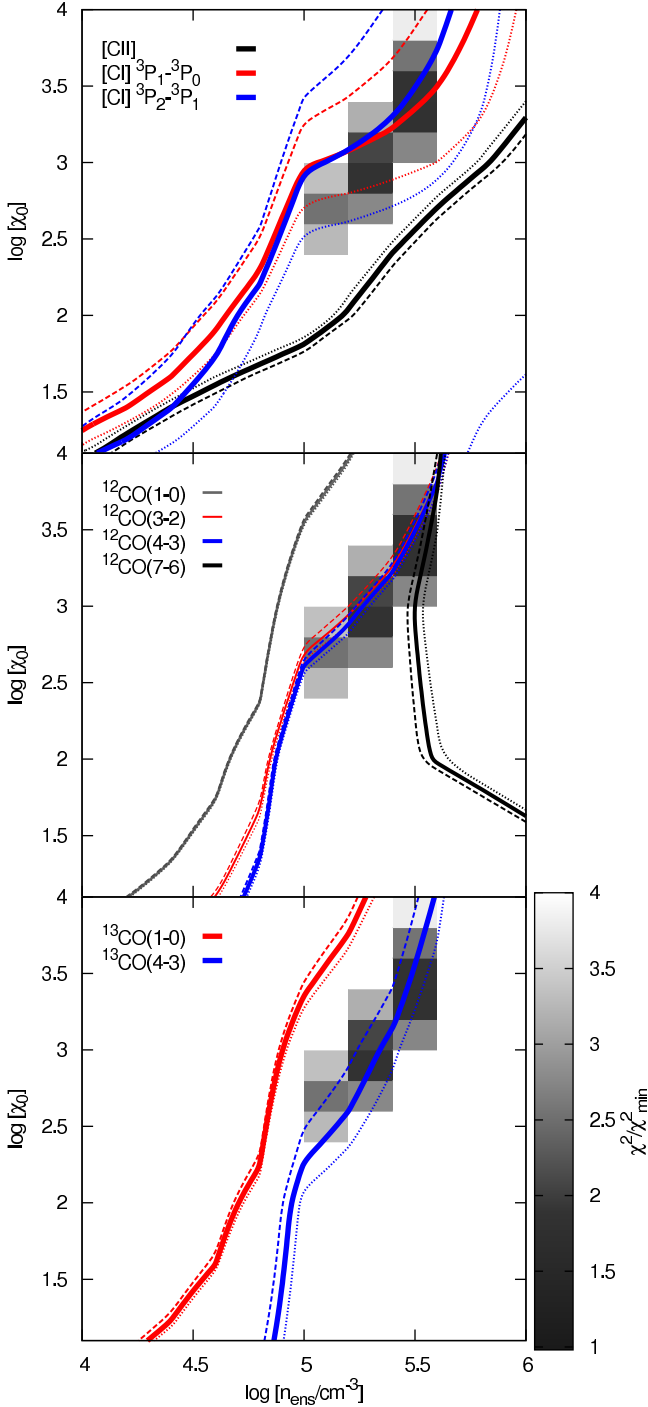}
     \caption{Comparison between the different observed lines with the
     predictions from the clumpy--PDR model for a given H$_2$ volume
     density and FUV radiation field.  In all panels, the solid lines
     represent the observed values while the dashed and dotted lines
     represent their lower and upper error bars, respectively. The
     gray-scale represents the value of $\chi^2$ relative to its
     minimum value, $\chi^2_{min}$. }\label{fig:pdr_fit}
\end{figure}


 The best--fit model results suggest an average ensemble total H
volume density $n_{\rm ens} = 3.3\times10^5$\,cm$^{-3}$, an ensemble
mass $M_{\rm ens} = 1.2\times10^5$ M$_{\odot}$.  The minimum and
maximum masses of the clump ensemble are $M_{\rm min} =
1\times10^{-3}$\,M$_\odot$ and $M_{\rm max} = 0.1$\,M$_\odot$.  Note
that $n_{\rm ens}$ corresponds to the average total H volume density
(i.e. $n_{\rm H}$+2$n_{\rm H_2}$). Ignoring the contribution from
atomic H, the average ensemble H$_2$ volume density is about
$6\times10^4$\,cm$^{-3}$, consistent with that derived with the
excitation analysis in Section~\ref{sec:escape-prob-radi}. The
ensemble mass is also consistent with that derived for the whole
extent of the 30\,Dor--10 cloud in Section~\ref{sec:column_densities}.
A comparison between the observations, the predictions from the PDR
model, and the results of the excitation analysis
(Section~\ref{sec:escape-prob-radi}) is listed in
Table~\ref{tab:model_comparison}.

 The area filling factor of the clump ensemble is defined as the sum
of all clump projected areas over the cloud solid angle (assumed to be
90$\arcsec$). Similarly, the volume filling factor is defined as the
sum of all clump volumes over the volume of a sphere with 90$\arcsec$
projected diameter. The model fit gives an area filling factor of 2.94
and, assuming a distance of 50\,kpc, a volume filling factor of
0.001. These area and volume filling factors imply that the model
requires many low--mass clumps to reproduce the observed line
intensities. As discussed by e.g.  \citet{Kramer2008}, these low--mass
clumps would require very high external thermal pressures to be
confined and might be considered as transient features of the
turbulent gas and should evaporate on short timescales.  Note that the
area filling factor defined here is not comparable with the beam
filling factor derived in Section~\ref{sec:co-emission}, because the
PDR model does not predict whether these small clumps are clustered or
dispersed within the cloud.

 In Figure~\ref{fig:pdr_fit} we show the dependence of the different
observed transitions on different H$_2$ volume densities and FUV
radiation fields. Solutions in the range $n_{\rm
ens}=1\times10^5-1\times10^{5.6}$ cm$^{-3}$ and $\chi_0 = 400-6300$
are within three times the minimum $\chi^2_{min}=59$. Note that this
high value of $\chi^2$ is dominated by the inability of the PDR model
to reproduce the [C\,{\sc ii}] emission. If we exclude the [C\,{\sc
ii}] line in the calculation of $\chi^2$, we obtain a much lower
$\chi^2_{min}$ of 0.01, without significant changes in the constrained
physical conditions.

\begin{figure}[h]
  \centering
  \includegraphics[width=0.465\textwidth,angle=0]
{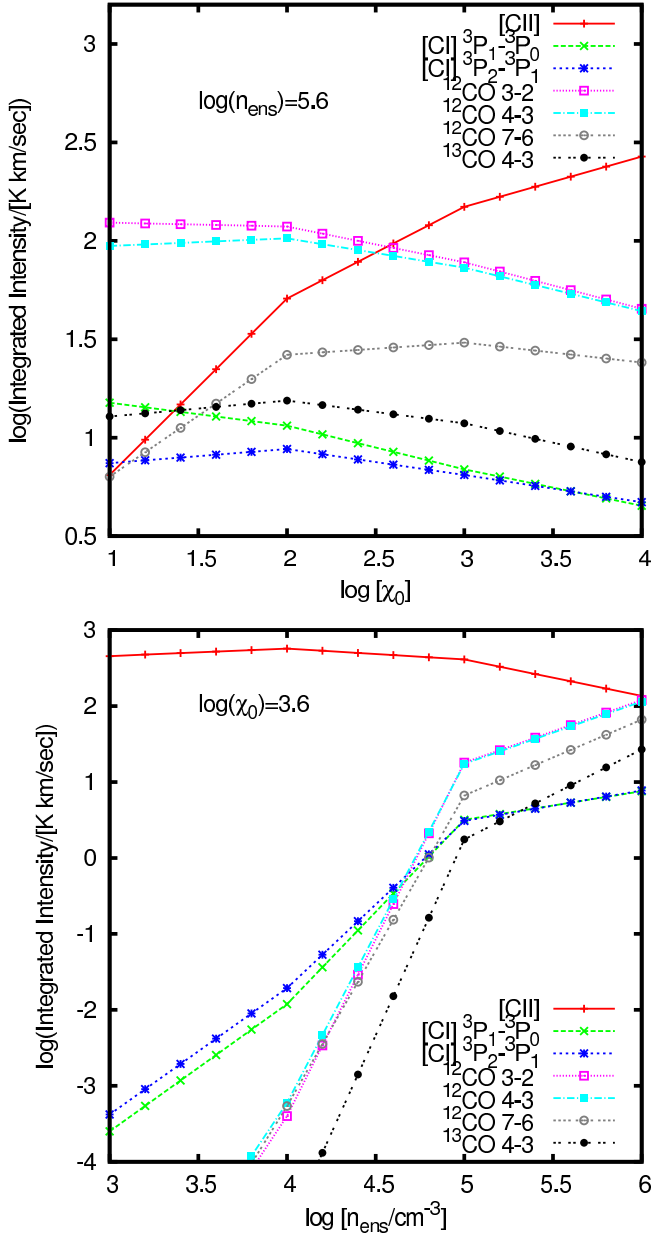}
     \caption{Model--predicted integrated intensities the spectral
lines used to constrain the PDR model as a function of $\chi_0$ for a
fixed value of $n_{\rm ens}$ ({\it upper panel}) and of $n_{\rm ens}$
for a fixed value of $\chi_0$ ({\it lower panel}).}
\label{fig:cuts}
\end{figure}

To illustrate the importance of the observed line intensities in
determining the value of $\chi_0$ and $n_{\rm ens}$, we show in
Figure~\ref{fig:cuts} the predicted model line intensities as a
function of $n_{\rm ens}$ for a fixed value of $\chi_0$=3980, and as a
function of $\chi_0$ for a fixed value of $n_{\rm
ens}$=3.9$\times10^{5}$\,cm$^{-3}$.  The fixed values of $\chi_0$ and
$n_{\rm ens}$ were selected to be the closest point in the model grid
to the solution found using the simulated annealing technique.  The
intensity of the mid--$J$ $^{12}$CO and $^{13}$CO are very sensitive
to $n_{\rm ens}$ and drive the value of this parameter.  The [C\,{\sc
i}] and mid--$J$ CO lines are sensitive to the FUV radiation field but
to lesser extent than to $n_{\rm ens}$. Consequently, as also seen in
Figure~\ref{fig:pdr_fit}, the fit to the average ensemble volume
density is better than that for the FUV radiation field.  The
decreasing value of the [C\,{\sc i}] and mid--$J$ CO lines and the
increase of value of the [C\,{\sc ii}] intensity with $\chi_0$ is
related to the shift of the C$^+$/C$^0$/CO transition deeper into the
cloud as $\chi_0$ increases, reducing the column density of CO and
increasing column density of [C\,{\sc ii}].  As shown by
\citet{Cubick2008}, the emission from CO transitions with $J>5$ is
mostly arising from low--mass clumps, while lower--$J$ CO and [C\,{\sc
i}] transitions arise from clumps with larger masses. Therefore, the
observed $^{12}$CO $J=7 \to 6$ transition is very sensitive to the
lower mass limit and drives the value of this model parameter. For the
upper mass limit, the dependence on the observed lines is weaker, and
varying the upper mass limit to higher values does not have a
significant effect on the model fit.

In Figure~\ref{fig:pdr_fit} and Table~\ref{tab:model_comparison} (see
also Figure~\ref{fig:sled}), we can see that the best--fit model
agrees well with observations of mid$-J$ CO and [C\,{\sc i}] but it
overestimates the [C\,{\sc ii}], $^{12}$CO $J = 1 \to 0$, and
$^{13}$CO $J = 1 \to 0$ line intensities. To reproduce the lines
observed by NANTEN2, the model requires clumps with volume densities
that are much higher than the critical density of the [C\,{\sc ii}]
line ($5\times10^3$ cm$^{-3}$; see Table~\ref{tab:line_parameters}),
resulting in a predicted [C\,{\sc ii}] intensity that is higher than
observed. A model with a lower volume density might reproduce the
[C\,{\sc ii}] line better, but it would underestimate the mid$-J$ and
[C\,{\sc i}] transitions. Another complication is that, as shown in
Figure~\ref{fig:plot_all_spectra}, we can expect that the FUV field is
not uniform within 30\,Dor--10, because parts of the cloud are closer
to R136 than others, while the model assumes a uniform field.  Note
that we are only considering a single--component model in this
analysis. It might be possible to find a better fit to the lines if
more gas components were added, but more data (e.g. [O\,{\sc i}]
lines) would be needed to obtain a good constraint to the model.  Note
that, as mentioned above, the PDR model neglects any mutual
absorption/shielding between clumps. Since the $^{12}$CO $J = 1 \to 0$
is optically thick and $^{13}$CO $J = 1 \to 0$ is likely affected by
optical depth effects, radiation from background clumps should be
(partly) shielded by foreground clumps. Because the model ignores this
effect, it overestimates the intensity of these transitions. This
effect might also partly be the reason for the overestimation of the
[C\,{\sc ii}] line, because this line might be also affected by
optical depth effects between clumps.

Our determination of the strength of the FUV field agrees excellently
with the independent estimation presented by \citet{Poglitsch95} of
$\chi_0$ = 3100--3600.  The constraints on the H$_2$ volume density
and FUV radiation field for 30\,Dor are also consistent with previous
determinations based on PDR modeling
\citep{Pak1998,Kaufman99,Bolatto99,Roellig06}. A crucial test to the
suggested clumpy structure of the LMC's ISM will be high angular
resolution imaging of low-- and mid--$J$ CO and [C\,{\sc i}] transitions
with ALMA.



\section{Discussion}
\label{sec:Discussion}

\subsection{Comparison with LMC--N159W}
\label{sec:comparison-with-lmc}

The 30\,Dor--10 region is located at a distance of $\sim20$\,pc from
the center of the R136 star cluster, hence this region is exposed to
an extreme FUV radiation field. It is of interest to compare its
physical properties with other regions in the LMC that have more
moderate ambient radiation fields. Since there are no significant
spatial variations of the metallicity in the LMC \citep{Dufour1984},
this comparison will help us to understand the effect of the FUV
radiation field on low--metallicity environments.

In \citet{Pineda2008} we made a similar analysis of the same set of
spectral lines as analyzed here in the LMC--N159W region. We
constrained the H$_2$ volume density to be $\sim$$10^{4}$\,cm$^{-3}$,
a kinetic temperature of $\sim$80\,K and equal CO and C\,{\sc i}
column densities of $\sim$1.6$\times10^{17}$\,cm$^{-2}$.  The beam
filling in N159W is about 1/6.  Modeling of the photon-dominated
region suggests that the strength of the FUV field in N159W is about
$\chi_0$=220, while we find $\chi_0$ = 3160 in 30\,Dor--10
(Section~\ref{sec:pdr-model-analysis}).

The 30\,Dor--10 region is warmer than N159W, with a kinetic
temperature of about 160\,K and 80\,K, respectively.  This higher
temperature is indicated by the higher $^{12}$CO $J=7\to 6$ to $J =
4\to 3$ ratio in 30\,Dor--10. This warming can be understood because
the gas heating is likely dominated by photoelectric heating, which
depends on the strength of the FUV field \citep{BakesTielens98}.  The
higher temperatures in 30\,Dor--10 do not result in higher brightness
temperatures compared with N159W because of the reduced beam filling
(1/16).  The column/volume densities, however, are similar or somewhat
higher in 30\,Dor--10 compared with N159W, suggesting that CO clumps
in the former region are smaller. These smaller clumps can be
interpreted as the effect of the enhanced CO photo-destruction due to
the strong FUV radiation field.  This enhanced photo-destruction does
not result in a higher abundance of C\,{\sc i} relative to CO, becuase
the $N({\rm CO})/N$({\sc {C\,{\sc i}}}) column density ratio is even
higher in 30\,Dor--10, implying that the FUV field moves the
C$^{+}$/C/CO transition layer to locations deeper into the clumps, but
does not increase the thickness of this layer. Accordingly, most of
the carbon in 30\,Dor--10 should be in the form of C$^+$.  This effect
has been predicted by PDR models \citep[e.g.][]{Roellig06}.  Note that
we cannot eliminate other mechanisms that might destroy the molecular
cloud, such as stellar winds from massive stars. However, the N159W
should also be influenced by these effects.

We determined the column density of C$^+$ using the integrated
intensity observed by \citet{Boreiko91} of 7.7$\times
10^{-4}$\,erg\,cm$^{-2}$\,s$^{-1}$\,sr$^{-1}$. This quantity is
related to the C$^+$ column density, assuming a kinetic temperature
larger than 91\,K and a volume density higher than $n_{\rm
cr}=5\times10^{3}$\,cm$^{-3}$, by $N$(C$^+$)$=6.4\times10^{20}$
$I$([C\,{\sc ii}]) cm$^{-2}$($\,{\rm erg}\,{\rm cm}^{-2}\,{\rm
s}^{-1}\,{\rm sr}^{-1})^{-1}$ \citep{Crawford1985}.  With that we
obtain a column density of C$^+$ of 4.96$\times10^{17}$\,cm$^{-2}$ and
therefore the total carbon abundance in 30\,Dor--10 is distributed as
$N_\mathrm{C^+}:N_\mathrm{C}:N_\mathrm{CO}= $69\%:8\%:23\%. For
comparison, the distribution of carbon in the N159W region is
46\%:27\%:27\% while in the Galactic PDR DR21(OH) it is 3\%:10\%:87\%
\citep{Jakob2007}. As in N159W, the 30\,Dor--10 region has an
increased fraction of the gas--phase carbon in the form of C$^+$
compared with DR21(OH).

%

\begin{figure*}[]
  \centering
  \includegraphics[width=\textwidth,angle=0]
{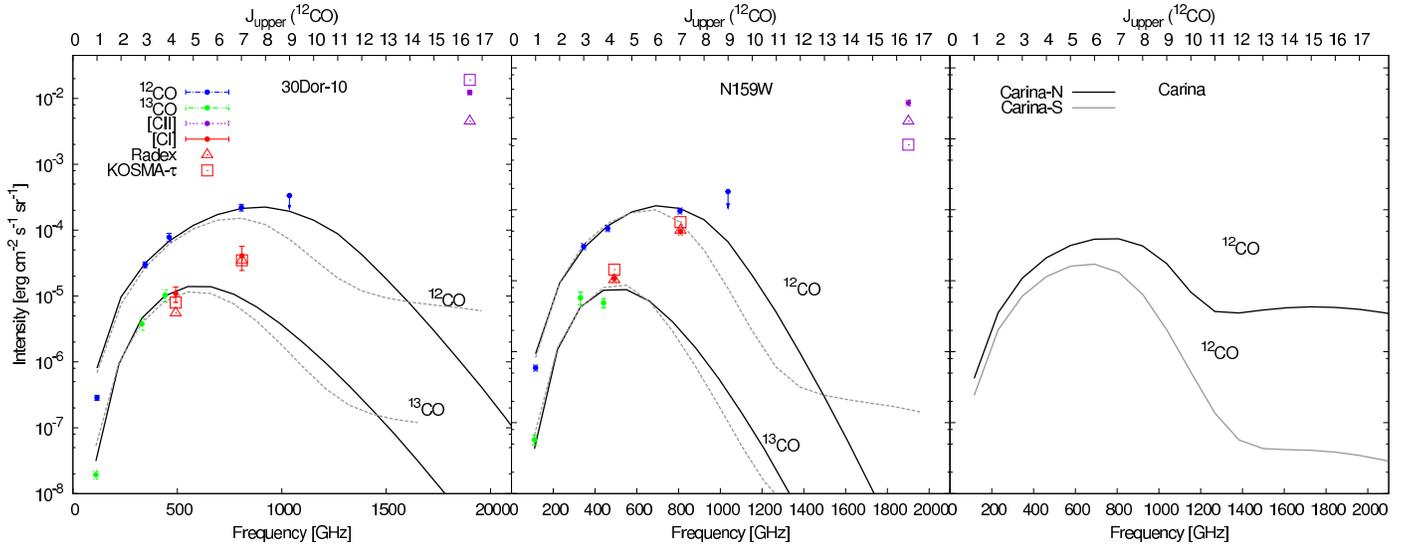}
     \caption{ Spectral--line energy distribution (SLED) for several
millimeter and submillimeter lines toward 30\,Dor--10 (left), N159W
(center), and Carina--N and --S (right).  In the left and central
panel, the solid lines represent the predictions for the $^{12}$CO and
$^{13}$CO SLED from the excitation analysis presented in
Section~\ref{sec:escape-prob-radi} for 30\,Dor--10 and in
\citet{Pineda2008} for N159W. The gray--dashed lines represent the
$^{12}$CO and $^{13}$CO SLED predicted by the clumpy PDR model
fits. The excitation analysis and PDR model predictions for the
[C\,{\sc i}] and [C\,{\sc ii}] lines are represented by triangles and
boxes, respectively. In the right panel we show the $^{12}$CO SLED for
the Carina--N and Carina--S regions presented by \citet{Kramer2008}
using the same clumpy PDR model as used in this paper. The physical
conditions derived from the PDR model are similar for 30\,Dor--10 and
Carina--N as well as for N159W and Carina--S, but they have different
metallicity (Carina $Z$=Z$_{\odot}$; LMC $Z$=0.4
Z$_{\odot}$). }\label{fig:sled}
\end{figure*}

\subsection{CO spectral--line  energy distribution}
\label{sec:co-line-spectral}

We combined our observed mid--$J$ $^{12}$CO observations with
previously observed transitions to derive the CO spectral--line energy
distribution (SLED) in 30\,Dor--10 and N159W.  We present the CO SLEDs
in Figure~\ref{fig:sled}, together with the predictions from our
excitation analyses and PDR modeling in both regions.  We include
observations of $^{12}$CO $J = 3 \to 2$ from \citet{Minamidani2007}
and upper limits to the intensity of the $J = 9 \to 8$ line from
\citet{Boreiko91}.  We corrected these line intensities, together with
that of the $J = 7 \to 6$ transition, to match the intensity
corresponding to a 38\arcsec\, beam (see
Section~\ref{sec:observ-results}).  Single--component models agree
well with the observed CO transitions. But the model for 30\,Dor--10
still predicts a line intensity of the $J = 1 \to 0$ line that is
about a factor of 3 higher than observed.  The $^{12}$CO SLED peaks at
$J= 8 \to 7$ for 30\,Dor--10 and $J= 6 \to 5$ for N159W.  The
different CO excitation in both regions reflects the difference in
their kinetic temperature, which in turn is due to the difference in
the FUV radiation field between both regions.

The CO SLED predicted by the clumpy PDR model for 30\,Dor--10 is
similar to that predicted by the excitation analysis for low--$J$
transitions, but differs significantly for high--$J$ transitions, with
the PDR model predicting much higher intensities for $J>6$.  For
$J>14$, however, the PDR model predicts almost constant intensities,
while the intensity predicted by the excitation model drops. The
clumpy PDR model predicts a peak of the CO SLED at a lower $^{12}$CO
transition ($J = 6 \to 5$) compared with the prediction from the
excitation analysis ($J = 8 \to 7$).  For N159W the clumpy PDR model
CO SLED also predicts lower intensities for high--$J$ transitions but
the difference is smaller than for 30\,Dor--10. The peak of the CO
SLED still coincides with that predicted by the excitation analysis
($^{12}$CO $J = 6 \to 5$).  More observations of high--$J$ CO
transitions with e.g. Herschel or SOFIA in 30\,Dor--10 and N159W are
required to test the predictions of the PDR modeling and excitation
analysis for high--$J$ CO transitions.

In Figure~\ref{fig:sled} we also show the CO SLED for two positions in
the Carina region, Carina--N and Carina--S, whose physical conditions
have been constrained with the KOSMA$-\tau$ PDR model by
\citet{Kramer2008} using the same set of line as presented here. The
constrained values of ($\chi_0$, $n_{\rm ens}$) are (3100,
$2\times10^{5}$ cm$^{-3}$) for Carina--N and (310, $2\times10^{5}$
cm$^{-3}$) for Carina--S. Therefore, both Carina--N and Carina--S have
similar physical conditions compared with 30\,Dor-10 and N159W,
respectively, but solar metallicity. The metallicity, however, seems
not to affect the shape of the CO SLED because for both pairs of
sources the CO spectral energy distribution is consistent within the
small differences in the physical conditions and uncertainties of the
models.

 The total CO luminosity ($L_{\rm CO}$) predicted by the radiative
   transfer code in an area defined by the 38\arcsec\,beam of our
   NANTEN2 observations (corresponding to 9.5\,pc at a distance of
   50\,kpc) is 1.1$\times$10$^{36}$ erg s$^{-1}$ for 30\,Dor--10 and
   4.3$\times$10$^{35}$ erg s$^{-1}$ for N159W. The clumpy PDR model
   predicts a somewhat lower $L_{\rm CO}$ of 8.2$\times$10$^{35}$ erg
   s$^{-1}$ for 30\,Dor--10 and 2.5$\times$10$^{35}$ erg s$^{-1}$ for
   N159W.  From the observations, we also estimated the total [C\,{\sc
   ii}] and [C\,{\sc i}] luminosity (ignoring the much weaker [C\,{\sc
   i}] $^3$P$_2 \to ^3$P$_0$ transition) of 9.9$\times$10$^{36}$ and
   4.2$\times$10$^{34}$ erg s$^{-1}$, respectively for 30\,Dor--10 and
   2.6$\times$10$^{36}$ and 4.9$\times$10$^{34}$ erg s$^{-1}$,
   respectively, for N159W.  The relative fraction of the luminosity
   of these species is $L_\mathrm{[C\,{\sc II}]}:L_\mathrm{[C\,{\sc
   I}]}:L_\mathrm{CO}$=90\%:0.4\%:9.6\% for 30\,Dor--10 and
   84.4\%:1.6\%:14 for N159W, when using $L_{\rm CO}$ predicted by the
   radiative transfer code, and 92\%:0.4\%:7.6\% for 30\,Dor--10 and
   90\%:1.7\%:8.3\% for N159W, when using $L_{\rm CO}$ predicted by
   the clumpy PDR model.  We also estimates a total [O\,{\sc i}]
   luminosity of 7.4$\times$10$^{36}$ erg s$^{-1}$ from observations
   of the [O\,{\sc i}] 63$\mu$m and 146$\mu$m lines by
   \citet{Poglitsch95} in 30\,Dor. These estimated luminosities
   suggest that [C\,{\sc ii}] and [O\,{\sc i}] contribute equally to
   the gas cooling in 30Dor--10. This significant contribution of
   [O\,{\sc i}] to the line cooling confirms the warm and dense gas
   conditions derived with the excitation analysis and PDR modeling.
   Note that we cannot rule out the possibility of an additional
   component powering even higher CO transitions ($J>10$) in
   30\,Dor--10 and N159W.  Therefore, our estimate of the total CO
   luminosity might be a lower limit. As mentioned above more
   observations will allow us to determine the contribution of
   high$-J$ CO transitions to the total CO luminosity.

\section{Summary and conclusions}
\label{sec:summary-conclusions}
We derived the physical conditions of the line-emitting gas in the
30\,Dor--10 region in the LMC. We compared an excitation/radiative
transfer code and a PDR model with NANTEN2 observations of the
$^{12}$CO $J = 4 \to 3$, $J = 7 \to 6$, and $^{13}$CO $J = 4 \to 3$
rotational and [C\,{\sc i}] $^3$P$_1-^3$P$_0$ and $^3$P$_2-^3$P$_1$
fine-structure transitions. Our results can be summarized as follows:

\begin{itemize}
  
\item The analysis of the excitation conditions for both the CO submm-lines
  and the [C\,{\sc i}] fine structure lines shows temperatures of about $T =
  160$\,K and densities of about $n_{\rm H_2}= 10^{4}$ cm$^{-3}$ for the emitting gas.

\item We found that 30\,Dor--10 is warmer and has a lower beam filling
factor compared to N159W. This difference might be the result of the
FUV radiation field heating the gas and photodissociating CO
molecules. This, however, does not result in an enhanced C abundance
relative to CO: we obtained a $N({\rm CO})/N$({\sc {C\,{\sc i}}})
ratio of 3 in 30\,Dor--10, which is higher than that found in N159W of
$N({\rm CO})/N$({\sc {C\,{\sc i}}}) equal to unity.

\item We derived the CO spectral--line energy distribution of
30\,Dor--10 and N159W. Considering the excitation conditions
constrained using line ratios, we found that this distribution peaks
at $^{12}$CO $J = 8\to7$ for 30\,Dor--10 and at $J=6\to5$ in
N159W. The clumpy PDR model, however, predicts the peak of the CO
spectral--line energy distribution at $J = 6\to 5$ for both
30\,Dor--10 and N159W.


 \item We compared our observations with the results of a clumpy PDR
model. The model that best reproduces the observed absolute integrated
intensities has an average ensemble H density of $n_{\rm ens} \sim
10^5$\,cm$^{-3}$, a total mass of the ensemble, $M_{\rm ens} \sim
10^5$ M$_{\odot}$, and a strength of the FUV field of $\chi_0 \sim 3
100$. The constraints on the H$_2$ volume density and the strength FUV
radiation field agree well with independent determinations and
previous PDR modeling results.

\end{itemize}

The 30\,Dorarus region is the closest example of vigorous
star--formation taking place in a low--metallicity environment, which
is thought to be common in the early universe.  Our results suggest
that the star--forming ISM in 30\,Dor--10, and perhaps in starburst
galaxies in the early universe, is dense, and warm, and clumpy. The
clumpy structure of the 30\,Dor-10 gas is a result of the extreme FUV
environment to which the low--metallicity gas is exposed. Our results
also highlight the importance of ionized carbon in tracing the total
gas mass and, together with atomic oxygen, in regulating the thermal
balance of the gas. Future observations of fine--structure transitions
of [C\,{\sc ii}], [O\,{\sc i}], [N\,{\sc ii}], and [C\,{\sc i}] and
rotational transitions of CO in the local Universe are essential for
our understanding of star--formation in low--metallicity environments
and are an important tool for the interpretation of observations of
high--redshift galaxies with ALMA.



\begin{acknowledgements} 
%
This research
 was conducted at the Jet Propulsion Laboratory, California Institute
 of Technology under contract with the National Aeronautics and Space
 Administration. \copyright 2011. All rights reserved. 
We would like to thank William Langer and Paul Goldsmith for careful
 reading of the manuscript and enlightening discussions. 
  The NANTEN2 project (southern submillimeter observatory consisting of a
  4-meter telescope) is based on a mutual agreement between Nagoya University
  and The University of Chile and includes member universities from six
  countries, Australia, Republic of Chile, Federal Republic of Germany, Japan,
  Republic of Korea, and Swiss Confederation.
M.R. wishes to acknowledge support from FONDECYT(CHILE) grant 1080335.
\end{acknowledgements}   

\bibliographystyle{aa} \bibliography{papers}

\end{document}